\documentclass[aps,preprint,onecolumn,superscriptaddress,showpacs,floatfix]{revtex4-1}
\usepackage{amsmath}
\usepackage[pdftex]{graphicx} 
\newcommand{\be}{\begin{equation}}
\newcommand{\ee}{\end{equation}}
\newcommand{\bc}{\begin{center}}
\newcommand{\ec}{\end{center}}
\newcommand{\bi}{\begin{itemize}}
\newcommand{\ei}{\end{itemize}}
\newcommand{\ba}{\begin{eqnarray}}
\newcommand{\ea}{\end{eqnarray}}

\newcommand{\ignore}[1]{}

\begin{document}

\title{Electrophysiological correlates of non-stationary BOLD functional connectivity fluctuations}
\author{Enzo Tagliazucchi}\affiliation{Neurology Department and Brain Imaging Center, Goethe University, Frankfurt am Main, Germany.}
\author{Frederic von Wegner}\affiliation{Neurology Department and Brain Imaging Center, Goethe University, Frankfurt am Main, Germany.}
\author{Astrid Morzelewski}\affiliation{Neurology Department and Brain Imaging Center, Goethe University, Frankfurt am Main, Germany.}
\author{Verena Brodbeck}\affiliation{Neurology Department and Brain Imaging Center, Goethe University, Frankfurt am Main, Germany.}
\author{Helmut Laufs}\affiliation{Neurology Department and Brain Imaging Center, Goethe University, Frankfurt am Main, Germany.}

\date{\today}

\begin{abstract}

Spontaneous fluctuations of the BOLD (Blood Oxygen Level-Dependent) signal, measured with fMRI
(functional Magnetic Resonance Imaging), display a rich and neurobiologically relevant functional connectivity structure. 
This structure is usually revealed using time averaging methods, which prevent the detection of functional connectivity changes over time.
It has been recently shown that functional connectivity  fluctuates over time, with brain regions transiently coupling and 
decoupling during the duration of a typical resting state fMRI experiment. In this work we studied the electrophysiological correlates
of the aforementioned BOLD functional connectivity fluctuations, by means of  long ($\approx$ 50 min) joint electroencephalographic (EEG)
and fMRI recordings obtained from two independent populations: 15 awake subjects and 13 subjects undergoing vigilance transitions to light sleep.

We identified  widespread
positive and negative correlations between EEG spectral power and fMRI BOLD connectivity fluctuations in  a network of 90 cortical and subcortical 
regions. In particular, increased
alpha (8-12 Hz) and beta (15-30 Hz) power were related to decreased functional connectivity, whereas gamma (30-60 Hz)
power correlated positively with BOLD connectivity between specific brain regions.
Furthermore, these patterns were altered for subjects undergoing vigilance changes, with an involvement of the slow delta (0.4 - 4 Hz)
band in localized positive correlations. Finally, graph theoretical indices of network structure also exhibited sharp changes over time, with average path
length correlating positively with alpha power extracted from central and frontal electrodes.

Our results strongly suggest that non-stationary BOLD functional connectivity has a neurophysiological
origin. Positive correlations of BOLD connectivity with gamma can be interpreted as increased average binding over relatively long
periods of time, possibly due to spontaneous cognition occurring during rest. Negative correlations with alpha suggest functional inhibition
of local and long-range connectivity, associated with an idling state of the brain.

\end{abstract}

\maketitle

\section{Introduction}

The discovery of non-invasive brain imaging with functional Magnetic Resonance Imaging (fMRI) by means of the Blood Oxygen Level-Dependent (BOLD)
contrast has greatly improved our understanding of the living human brain. The observation of positive correlations between BOLD signals from
 anatomically distant brain areas \cite{biswal} (left and right motor cortex) provided an invaluable tool to characterize large-scale functional
connectivity. Remarkably, these correlations were observed during rest (i.e. when subjects were not being stimulated nor performing any task),
thus contributing to discard ongoing spontaneous BOLD fluctuations as experimental noise. The neurobiological relevance
of resting state functional connectivity is manifest by further studies uncovering synchronized spontaneous activity in systems co-activated during
task performance, termed Resting State Networks (RSN) \cite{greicius, beckmann2005, damoiseaux, fox2007,smith}. Multimodal EEG-fMRI studies demonstrated 
positive and negative correlations between spontaneous BOLD fluctuations in specific RSN
and spectral power in different EEG frequency bands \cite{laufs2003a, mantini}. Finally, functional connectivity between all areas of cortical and
subcortical whole-brain parcellations has been extensively studied using graph theoretical methods. For these analyses, each anatomical region is represented
as a node in a network, and two nodes are connected if the corresponding BOLD signals are highly correlated \cite{sporns, rubinov, bullmore}. The resulting
 functional connectivity networks
are small-world \cite{eguiluz, achard} (networks having a rich local connectivity structure and long distance
shortcuts) and scale-free \cite{eguiluz, vandenheuvel} (networks whose degree -or first neighbors- distribution follows a power law, and hence
 is scale invariant), properties characteristic
of systems  resistant to localized damage and efficient at large-scale information processing \cite{bullmore2}.

It must be noted that all aforementioned analyses, from simple temporal correlations \cite{greicius}
to more sophisticated model-free methods applied to identify RSN, such as Independent Component Analysis (ICA) \cite{beckmann2004}, yield results averaged over time.
Therefore, they assume that functional connectivity  is stationary and are unable to capture 
changes in the coordination between brain areas over time. However, recent work strongly suggests that the assumption of stationarity does not
 hold for large-scale BOLD functional connectivity. It has been shown that BOLD
time series can be decomposed into a discrete point process \cite{enzo1,enzo2,enzo3,petridou}, either by thresholding or deconvolution, providing 
 additional information about instantaneous synchronization between brain regions.
Time-frequency coherence analysis based on the wavelet transform was performed on BOLD time series, focusing on the posterior cingulate cortex (PCC) \cite{chang}.
This work revealed variable coherence and phase between the PCC and an anticorrelated network of brain regions, at time scales ranging from
seconds to minutes. Sliding window analyses also demonstrated similar variability over time in the coordination of anatomically distant cortical 
regions \cite{hutchinson}, not only in awake humans but also in anesthetized monkeys. Importantly, this eliminates spontaneous conscious cognition as
 the only source of 
non-stationary functional connectivity. Using similar methods, dynamical connectivity between the PCC and the rest of the brain was shown to have
a periodical structure \cite{handwerker}. Transient synchronization between brain regions was demonstrated for magnetoencephalography (MEG)
sources \cite{depasquale}, with complete RSN being revealed only when signal non-stationarities were taken into account.
Further evidence for dynamically changing large-scale coupling can be found in computational 
simulations of neural synchronization with realistic underlying structural connectivity \cite{honey}. In light of these results, the assumption of
stationary large-scale connectivity has to be revisited.

Transient synchronization between signals of electrophysiological origin has long been related to the performance of many cognitive operations \cite{fries1,fries2}.
Gamma band synchronization has been postulated as a fundamental process subserving an elementary operation of cortical computation \cite{fries3} and
as a plausible neurophysiological solution for the binding problem \cite{malsburg, treisman}.
On the other hand, synchronized activity in alpha and peri-central (or rolandic) beta bands was proposed to reflect idling \cite{pfurtscheller0}
 and inhibition \cite{klimesch, jensen}
of activity in task-irrelvant regions. A recent study \cite{scheeringa} showed that decreased BOLD connectivity of primary visual areas is linked to increased occipital
alpha power, interpreting this finding as a consequence of enhanced functional inhibition during periods of high alpha. In this work we systematically study
the relationship between whole-brain BOLD connectivity and power fluctuations in different frequency ranges.
Considering these previous experimental results, we hypothesize that band-specific electrophysiological spectral
 changes will parallel fluctuations in BOLD connectivity and that increased 
power in fast  EEG frequency bands (such as gamma) will correlate with an increase of BOLD connectivity, 
whereas slower frequencies (such as alpha) will not be expected contributors to the onset of transient,
long-range BOLD coordination. We further hypothesize that the observed changes in functional connectivity over time will result in changes of 
graph theoretical metrics (which are also usually assumed to be stationary over time), which might also be correlated with power fluctuations
 in specific frequency bands. We put this hypothesis to test by studying
two different populations undergoing joint EEG-fMRI recordings: a group of  awake subjects and a group of subjects undergoing vigilance transitions
to light sleep or drowsiness, in order to study possible correlations between BOLD functional connectivity changes and the slow frequencies heralding the onset of
sleep.

\section{Materials and methods}

\subsection{EEG-fMRI acquisition and artifact correction}

EEG via a cap (modified BrainCapMR, Easycap, Herrsching, Germany) was recorded continuously during fMRI acquisition (1505 volumes of T2*-weighted 
echo planar
 images, TR/TE = 2080 ms/30 ms, matrix 64 $\times$ 64, voxel size 3 $\times$ 3 $\times$ 2 mm$^3$, distance factor 50\%; FOV 192 mm$^2$) at 3 T 
(Siemens Trio,
 Erlangen, Germany) with an optimized polysomnographic setting (chin and tibial EMG, ECG, EOG recorded bipolarly [sampling rate 5 kHz, low pass 
filter 1 kHz],
 30 EEG channels recorded with FCz as the reference [sampling rate 5 kHz, low pass filter 250 Hz], and pulse oxymetry, respiration recorded 
via sensors
 from the Trio [sampling rate ~ 50 Hz]) and MR scanner compatible devices (BrainAmp MR+, BrainAmp ExG; Brain Products, Gilching, Germany). 
 
MRI and pulse artifact correction were performed based on the average artifact subtraction (AAS) method \cite{allen} as implemented in Vision Analyzer2
 (Brain Products, 
Germany) followed by objective (CBC parameters, Vision Analyzer) ICA-based rejection of residual artifact-laden components after AAS resulting 
in EEG with 
a sampling rate of 250 Hz. EEG was re-referenced to common average. Sleep stages were scored manually by an expert according to the AASM criteria \cite{AASM}.

\subsection{Subjects and datasets}

A total of 15 awake subjects were included in study (10 female, age 26.2 $\pm$ 6), together with an independent group of 13 subjects undergoing
vigilance transitions between wakefulness and light sleep (8 female, age 23.3 $\pm$ 3.4). 
Both groups were extracted from a larger dataset with the following inclusion 
criteria: for the first group, subjects did not show any period of sleep (as determined by AASM sleep scoring criteria).
For the second group, subjects showed only epochs of wakefulness and at least \%20 of light (N1) sleep.
In all cases, written informed consent and approval by the local ethics committee were obtained.

\subsection{fMRI pre-processing}

Using Statistical Parametric Mapping (SPM 8, http://www.fil.ion.ucl.ac.uk/spm) EPI data were realigned, normalized (MNI space) and spatially 
smoothed 
(Gaussian kernel, 8 mm$^3$ full width at half maximum). 
 Cardiac-, respiratory- 
(both estimated with the RETROICOR method \cite{glover}) and motion-induced 
noise were regressed out. fMRI data was bandpass filtered in the range 0.01-0.1 Hz using a 6th order Butterworth filter. 

\subsection{Time dependent correlation matrix}

To study the covariance of functional connectivity and EEG power in different frequency bands, an estimate of
how the former changes over time is necessary. For this purpose (following \cite{hutchinson,fraiman}) a sliding window analysis was employed, with a window length 
of $\approx$ 2 min (60 volumes). This 
window length was chosen because it is relatively short when compared to the length of the experiment ($\approx$ 50 min), while allowing good
functional connectivity estimates \cite{vandijk}. As a first step, the average BOLD signal was extracted from each one of the 90 cortical 
and subcortical regions defined in the AAL template \cite{tzourio} (information on all the regions is provided in Table 1).
 These time series are notated as $x_i, 1 \leq i \leq 90$. Next, the time dependent
correlation matrix was obtained as follows:

\begin{align}
 C_{ij}(t) = \frac{  \sum_{n=t}^{t+k} \left( x_i(n) - \frac{1}{k} \sum_{m=t}^{t+k} x_i(m) \right) \left( x_j(n) - \frac{1}{k} \sum_{m=t}^{t+k} x_j(m) \right) }{ \sqrt{\sum_{n=t}^{t+k} \left( x_i(n)- \frac{1}{k} \sum_{m=t}^{t+k} x_i(m)  \right)^2 } \sqrt{  \sum_{n=t}^{t+k} \left( x_j(n)- \frac{1}{k} \sum_{m=t}^{t+k} x_j(m)  \right)^2 }} \notag \\
=\frac{ \langle \left( x_i(t:t+k) - \langle x_i(t:t+k) \rangle \right) \left(  x_j(t:t+k) - \langle x_j(t:t+k) \rangle \right)  \rangle }{ \sigma \left(x_i(t:t+k) \right) \sigma \left(x_j(t:t+k) \right)} 
\end{align}

The second expression was simplified using a notation similar to MATLAB vector syntax, in which $x(n:m)$ represents the
portion of $x$ ranging from the n-th to the m-th entries. Thus, $C_{ij}(t)$ is the linear correlation between $x_i$ and $x_j$ during a window
of length $k$ starting from $t$. As mentioned above, $C_{ij}(t)$ was computed with $k=60$, which corresponds to approximately 2 min.

\subsection{EEG power, motion, cardiac and respiratory time courses}

\begin{figure}[htp]
\centering
\includegraphics[totalheight=0.8\textheight]{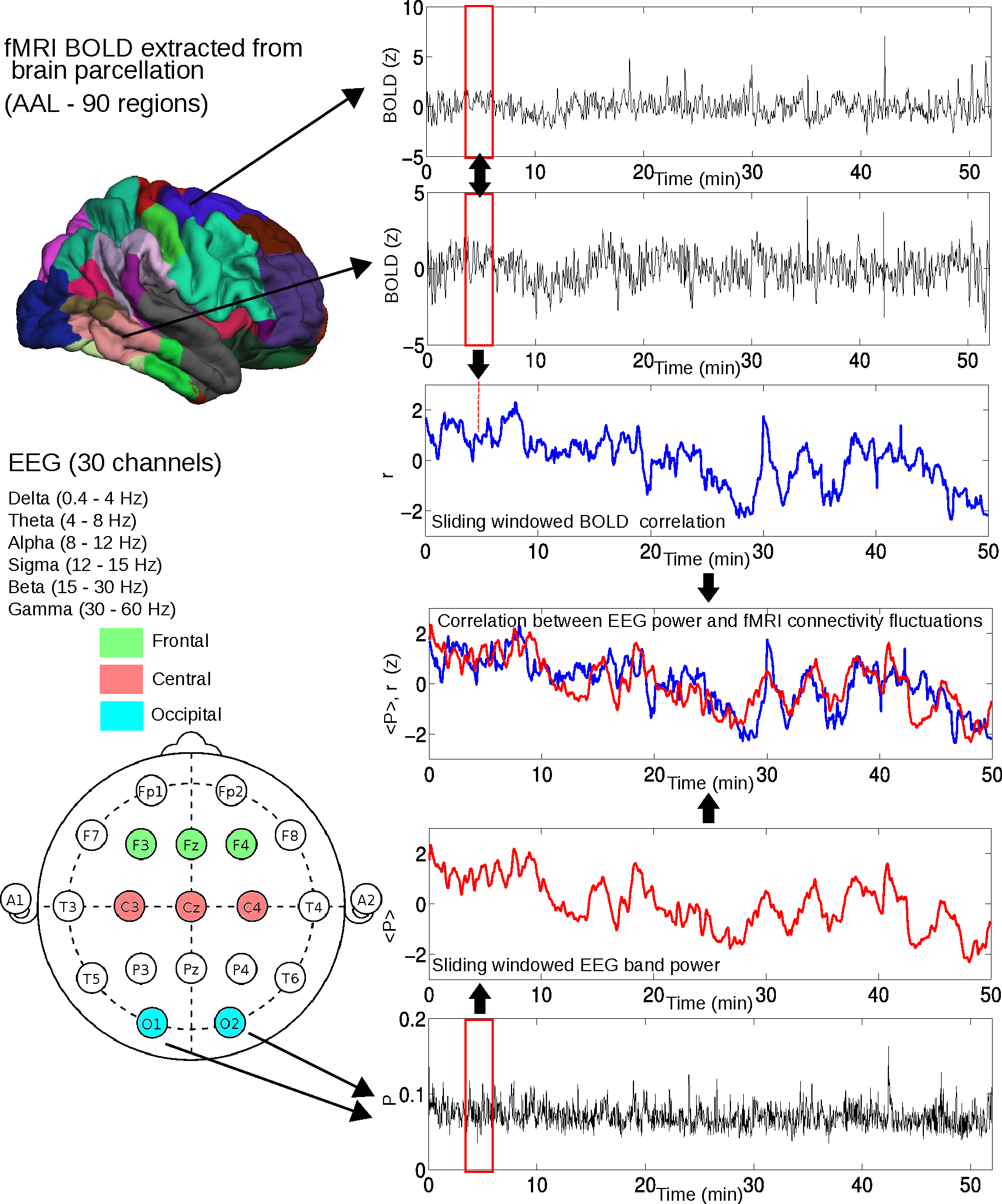}
\caption{Method used to compute BOLD connectivity
 fluctuations and correlations with EEG power fluctuations. For each pair of regions, average BOLD signals are extracted
 and correlated using a sliding window of 60 volumes ($\approx$ 2 min). This results in a connectivity estimate over time. A similar
 sliding window approach is used to obtain the average EEG power from different frequency bands (delta,theta,alpha,sigma,beta,gamma),
 averaged from different locations (frontal, central, occipital). As a final step, these EEG power fluctuations are correlated
 with BOLD connectivity for each pair of regions and correlations are tested for statistical significance (Student's t-test,
 FDR controlled for multiple comparisons)}.\label{fig1}
\end{figure}

Next, time courses for the variables to be correlated with BOLD connectivity were obtained. For this purpose, a sliding window 
average was applied, with the same window length ($\approx$ 2 min) used to construct the time dependent functional connectivity matrix ($C_{ij}(t)$). Given a 
time series $y$, the computation is as follows:

\begin{equation}
Y(t) = \frac{1}{k} \sum_{n=t}^{t+k} y(n) = \langle y(t:t+k) \rangle
\end{equation}

which gives the desired result when $k=60$. These sliding window averaged time courses were obtained for delta (0.4 - 4 Hz),
theta (4 - 8 Hz), alpha (8 - 12 Hz), sigma (12 - 15 Hz), beta (15 - 30 Hz) and gamma (30 - 60 Hz) power, averaged over frontal
(channels F1, Fz, F2), central (channels C1, Cz, C2) and occipital (channels O1, Oz, O2) EEG and also for the cardiac
and respiratory time series estimated with the RETROICOR method. The sliding window averaged time course for
the relative displacement with respect to an arbitrary volume was also obtained, computed as $D = \sqrt{D_x^2 + D_y^2 + D_z^2 }$, where
$D_x$ , $D_y$ and $D_z$ are the estimated displacements (after spatial realignment with SPM8) in the $x$, $y$ and $z$ axis, respectively.
For an overview of the data analysis, see Fig. \ref{fig1}.

\subsection{Correlation between fMRI connectivity fluctuations and electrophysiological time series}

Next, the time dependent BOLD functional connectivity between each pair of regions (as represented in $C_{ij}(t)$) was correlated 
with the sliding window averaged
time series of electrophysiological origin. Even though cardiac, respiratory and motion time series were regressed out of the signal at
the pre-processing stage, they were still kept as partial regressors in the analysis. Therefore, we obtained the partial correlation between 
$C_{ij}(t)$ for $1 \leq i \leq 90$, $1 \leq j \leq i-1$ and the different frequency bands from frontal, central and occipital channels:

\begin{equation}
R_C(X,Y) = \min_Z R(X,Y|Z)
\end{equation}

$R(X,Y|Z)$ is defined as follows,

\begin{equation}
R(X,Y|Z) = \frac{R(X,Y) - R(Y,Z)}{\sqrt{1 - R(X,Z)^2} \sqrt{1 - R(Y,Z)^2 } }
\end{equation}

in which $R(X,Y)$ is  the linear correlation between both variables, as in Eq. 1.

\subsection{Time dependent graph metrics}

The presence of correlations between the time development of common graph metrics associated with the global fMRI functional connectivity networks and
time courses of EEG power fluctuations in different frequency bands was then analyzed.
Graph metrics summarize topological information (i.e. not explicitly related to brain anatomy or geometry) about brain 
connectivity. For this purpose a graph representation of functional connectivity is needed, in which each \emph{node} represents a 
brain anatomical region and a \emph{link} between two nodes represents significant functional connectivity between the BOLD signals from the nodes.
To obtain this representation, the correlation matrix $C_{ij}(t)$ was thresholded to obtain the time dependent adjacency matrix \cite{bassett} $A_{ij}(t)$ as follows:

\begin{equation}
A_{ij}(t) =
\left\{
	\begin{array}{ll}
		0  & \mbox{if } C_{ij}(t) < \rho \\
		1 & \mbox{if } C_{ij}(t) \geq \rho
	\end{array}
\right.
\end{equation}

In the adjacency matrix $A_{ij}(t)$ a $1$ represents a link between nodes $i$ and $j$ at time $t$. 
The arbitrary parameter $\rho$ was chosen so that in all cases the resulting networks had a link density of 0.10, i.e. 10\% of the total number
of possible links in the networks were actually present. For each time step, a number of graph metrics using the MATLAB
Brain Connectivity Toolbox \cite{rubinov} was computed. Heuristic 
definitions are provided below (illustrations exemplifying the different metrics can be found in Fig. \ref{fig10}A, for a detailed review
see \cite{sporns, bullmore, rubinov}):

\begin{itemize}
 \item \emph{Clustering coefficient ($\gamma$)}. The clustering coefficient of a given node is the probability that two nodes which are connected to it, are 
also connected between them. The clustering coefficient is then computed as the average of the clustering coefficient of all individual nodes.
 \item \emph{Average path length ($\lambda$)}. The distance between node $i$ and node $j$ is the minimum number of links which have to be crossed
when going from $i$ to $j$. The average path length is the average of the distance between all possible pairs of nodes in the network.
 \item \emph{Betweeness ($\beta$)}. A  path between node $i$ and node $j$ is defined as the sequence of linked nodes which have to be visited to go from 
 $i$ to $j$. A minimum path between node $i$ and node $j$ is a path with a number of links equal to the distance between $i$ and $j$. The betweeness of a given node
in the network is defined as the number of minimum paths of which that node is a member. The betweeness of the network is computed as the average betweeness of all individual nodes.
 \item \emph{Small-worldness ($\sigma$)}. To compute the small-worldness coefficient, networks are first randomized, scrambling their links at random
with the constraint of a preserved connectivity distribution. Then, the clustering coefficient ($\gamma_{Rand}$) and the average path
length ($\lambda_{Rand}$) of the randomized networks are computed. The small-worldness coefficient is then obtained as $\sigma=\frac{\gamma^*}{\lambda^*}$,
where $\gamma^* = \frac{\gamma}{\gamma_{Rand} }$ and $\lambda^* = \frac{\lambda}{\lambda_{Rand} }$. A value of $\sigma > 1$ is  regarded
as indicator of small-world structure in the network \cite{humphries}.
\end{itemize}

After obtaining the time courses of $\gamma(t)$, $\lambda(t)$, $\beta(t)$ and $\sigma(t)$, the presence of correlations with
time courses of EEG power in the delta, theta, alpha, sigma, beta and gamma bands was studied (taking cardiac, respiratory and motion time series into account
as partial regressors).

\subsection{Statistical testing}

To test for statistical significance, all correlation values were  first transformed to z-scores using the Fisher transform,
given by $z = artanh(r)$. Then, Student's t-tests were performed with the null hypothesis of zero correlation. To correct for the multiple comparisons
performed the False Discovery Rate (FDR) method was used, with a rate of $q=0.05$. For the correlation between graph metrics and
EEG power time courses, the more conservative Bonferroni correction was applied with $n=6$ (freq. bands) $\times 3$ (number of channel groups) $\times 4$ (number of graph metrics) $=72$.

\section{Results}

\subsection{Spontaneous BOLD connectivity fluctuations}

We started by assesing the presence of temporal fluctuations in BOLD connectivity, a necessary first step to perform the correlation analysis with
EEG power fluctuations.

\begin{figure}[htp]
\centering
\includegraphics[totalheight=0.5\textheight]{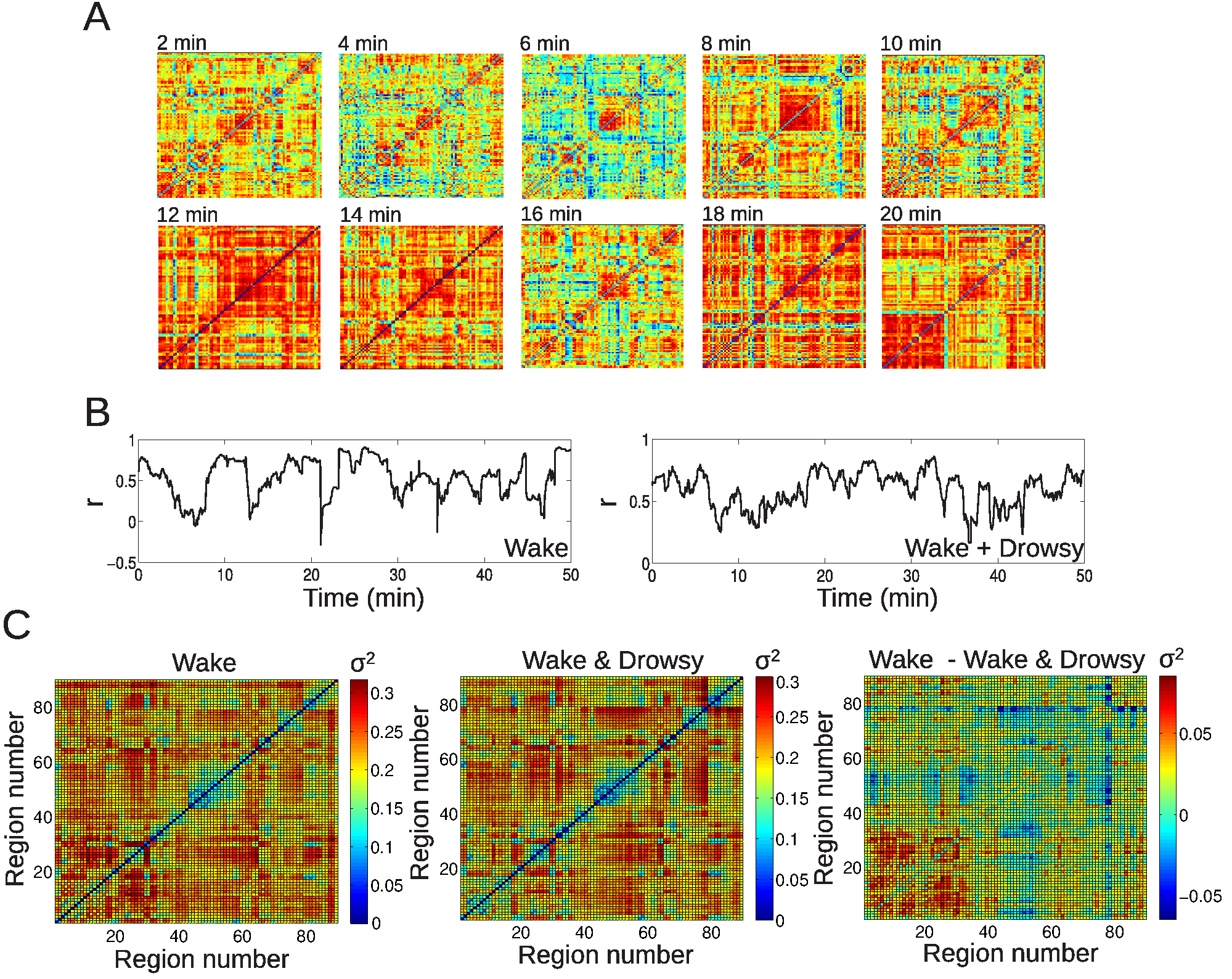}
\caption{Large-scale spontaneous BOLD functional connectivity fluctuations. A) BOLD correlation matrices for a single subject, in intervals of 2 min. B) Time series of
 inter-hemispheric thalamic connectivity for an awake subject and a subject undergoing vigilance transitions to light sleep. 
 C) Standard deviation of BOLD connectivity time series for each pair of regions, for both groups (wakefulness and wakefulness \& light sleep) and
 their difference.}\label{fig2}
\end{figure}

Functional connectivity between brain regions fluctuated widely over time, consistently with previous reports
by other research groups \cite{chang, hutchinson, handwerker}. An example of this dynamical functional connectivity is shown in Fig. 
\ref{fig2}A, in which the complete 
functional connectivity matrix is presented in intervals of 2 min for a single subject. It is clear, by simple visual inspection, that the connectivity matrix
fluctuates over time, with periods of overall decreased connectivity alternating with periods of globally increased connectivity (termed
\emph{hypersynchrony} in a previous study \cite{hutchinson}). As an example, in Fig. \ref{fig2}B the connectivity time courses between left and right thalamus are
 shown for the same
subject and also for a subject undergoing vigilance transitions to light (N1) sleep. Correlation between both BOLD signals clearly changes over time, 
alternating between a correlation close to the highest possible value ($r=1$) and a complete discoordination ($r=0$), in spite of strong inter-thalamic connectivity ($r>0.5$).
The extent of functional connectivity variation was quantified 
by taking the standard deviation (S.D.) of the connectivity time course between all pairs of regions. The average S.D. for both groups are shown in Fig. \ref{fig2}C,
together with their difference. 
``Blocks'' of lower S.D. can be observed along the diagonal of the matrix, indicating groups of (anatomically) neighboring regions with lower 
functional connectivity variability over time (for example, occipital regions, ranging from regions \#43-44 to \#55-56), 
while displaying a higher variability in their connections with the rest of the brain (off-diagonal entries). While there were no significative differences surviving
multiple comparison correction, there was a trend of higher S.D. in frontal connectivity for the group of awake subjects.

\subsection{Correlations with spontaneous EEG power fluctuations}

In the next stage of the analysis, we studied the correlations between EEG power and BOLD connectivity time courses, in order to identify EEG
frequency bands involved with connectivity changes over time.

\begin{figure}[htp]
\centering
\includegraphics[totalheight=0.8\textheight]{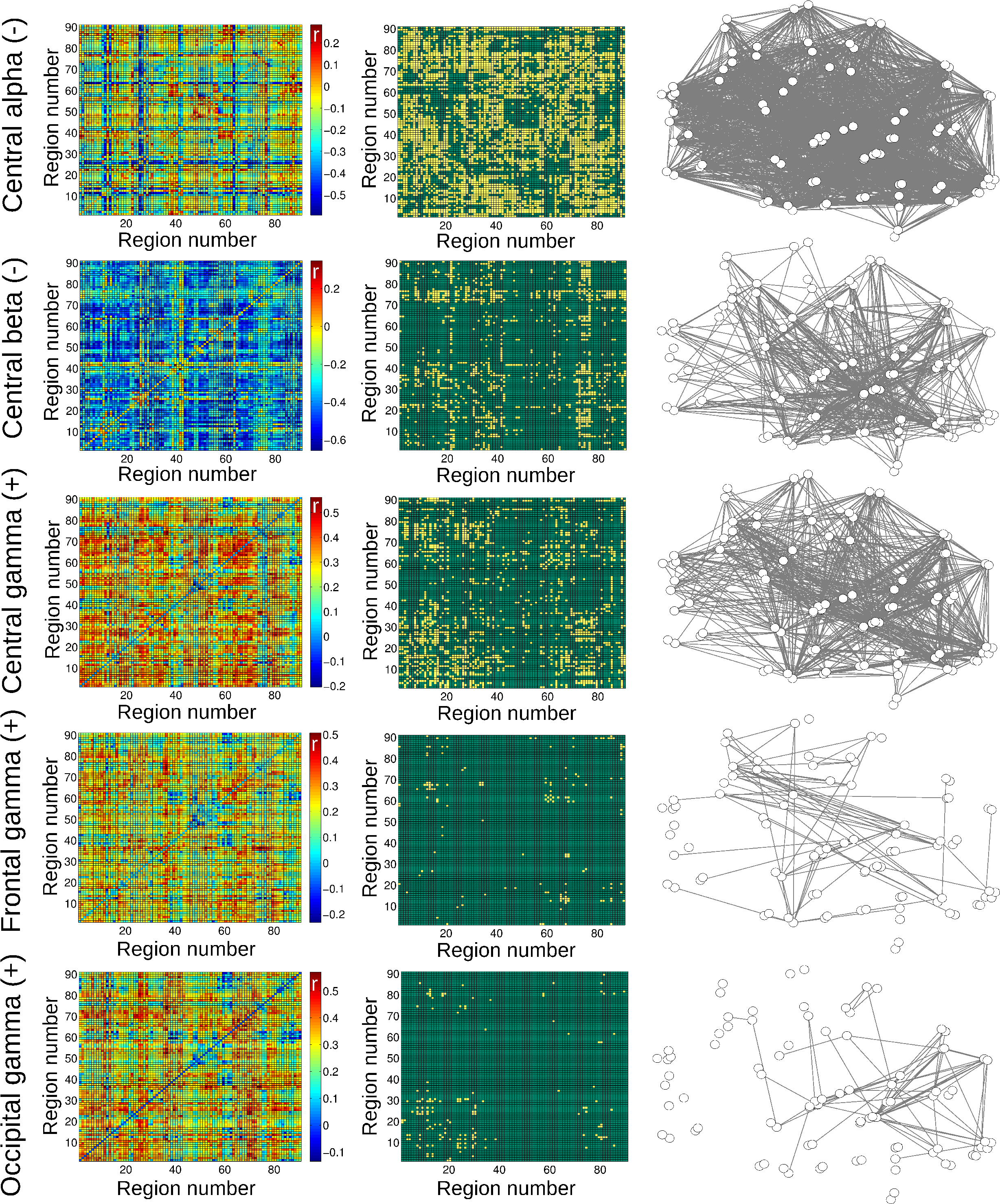}
\caption{Matrices of average
 correlation (left), significant correlations (center), and networks in anatomical space with links representing significant correlations (right). Correlations are between BOLD 
 functional connectivity and spontaneous EEG power fluctuations for all frequency bands and averaged from different anatomical locations. Results 
 are for the group of awake subjects.}\label{fig3}
\end{figure}

Results for the group of awake subjects are presented in Fig. \ref{fig3}. Widespread
negative correlations between BOLD connectivity fluctuations and central alpha and beta power were detected. Positive correlations were found with
central, frontal and occipital gamma power fluctuations, with correlations with central channels being more widespread than the others.

\begin{figure}[htp]
\centering
\includegraphics[totalheight=0.8\textheight]{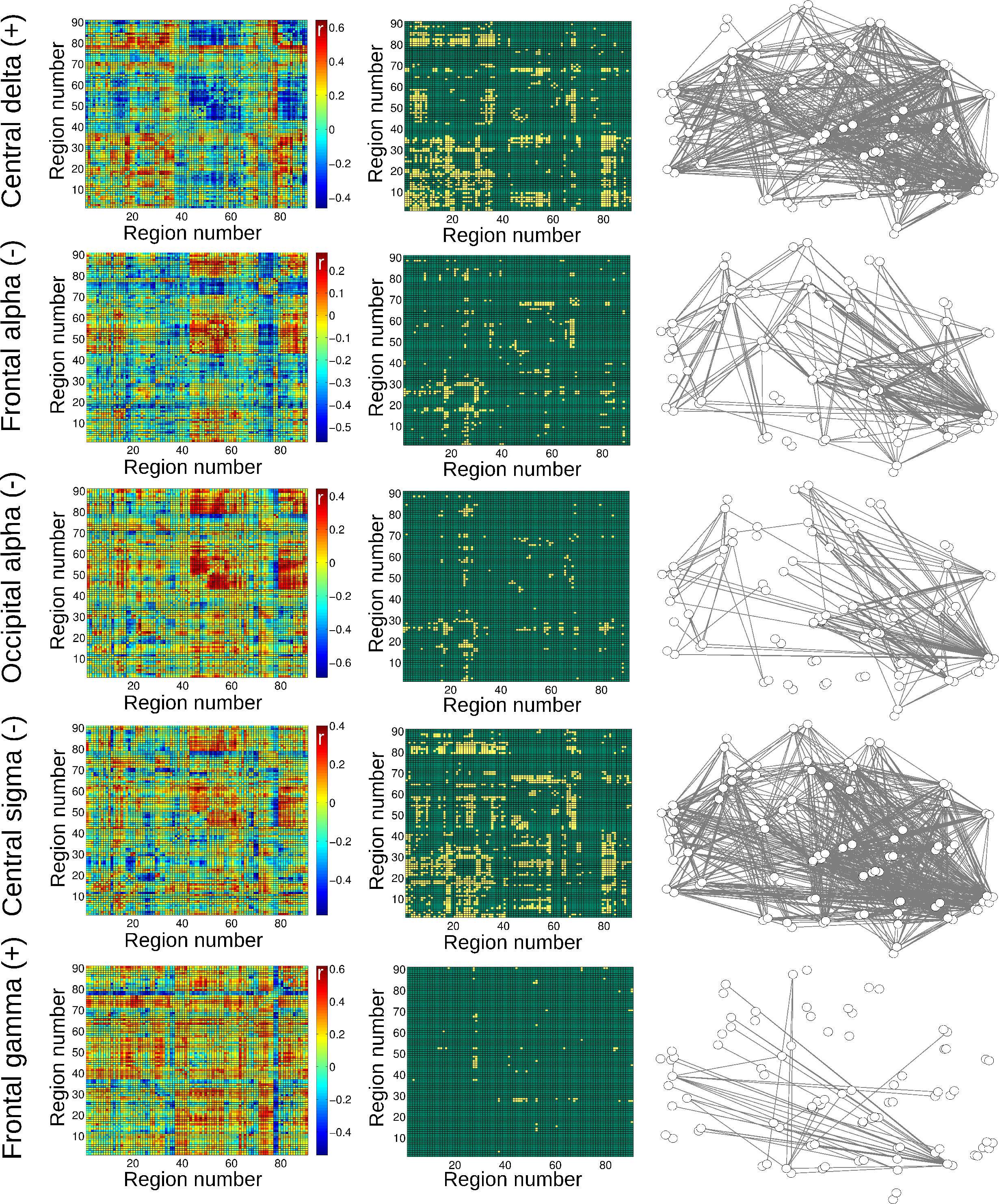}
\caption{Matrices of average
 correlation (left), significant correlations (center),  and networks in anatomical space with links representing significant correlations (right). Correlations are between BOLD 
 functional connectivity and spontaneous EEG power fluctuations for all frequency bands and averaged from different anatomical locations. Results 
 are for the group of subjects undergoing vigilance transitions to light sleep.}\label{fig4}
\end{figure}

Correlations for the group of subjects undergoing vigilance changes are presented in Fig. \ref{fig4}. Positive correlations were found with 
central delta power fluctuations, with a spatial emphasis in frontal regions and between frontal and temporal regions. Negative correlations 
with frontal and occipital alpha also affected predominantly frontal BOLD connectivity, whereas the central sigma band showed more 
distributed negative correlations. Finally, very few positive correlations with the frontal gamma time course were found.

\begin{figure}[htp]
\centering
\includegraphics[totalheight=0.3\textheight]{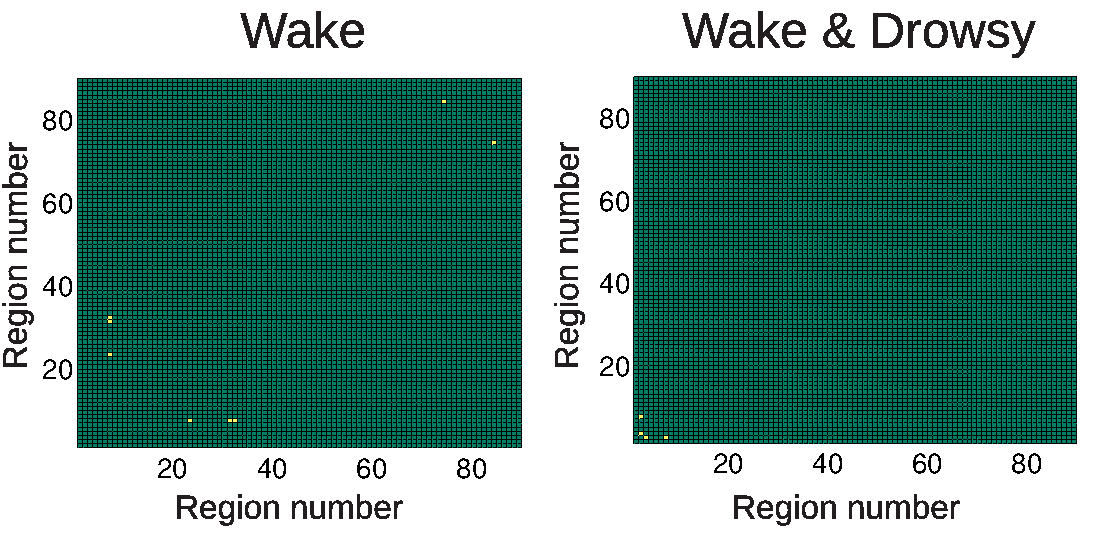}
\caption{Significant correlations between pairwise BOLD 
 connectivity fluctuations and respiratory time series, for the group of awake subjects (left) and the group of subjects undergoing
 vigilance transitions to light sleep (right).}\label{fig5}
\end{figure}

We directly tested whether functional connectivity changes correlated with cardiac, respiratory and motion time series. Negative
results were found for cardiac and motion time series, and very few positive correlations for the group of awake subjects (4 pairs of regions) and
for the subjects with fluctuating vigilance (2 pairs of regions). Results are presented in Fig. \ref{fig5}.

\subsection{Nodes with connectivities most influenced by EEG power fluctuations}

To extract information about the network of connections which are correlated with EEG power in specific frequency bands, we computed for each node the number
of connections with other nodes which were affected by spontaneous power fluctuations. This is equivalent to the degree (D) of each region or node,
if one defines a network whose links are the EEG power-modulated connections. 

\begin{figure}[htp]
\centering
\includegraphics[totalheight=0.8\textheight]{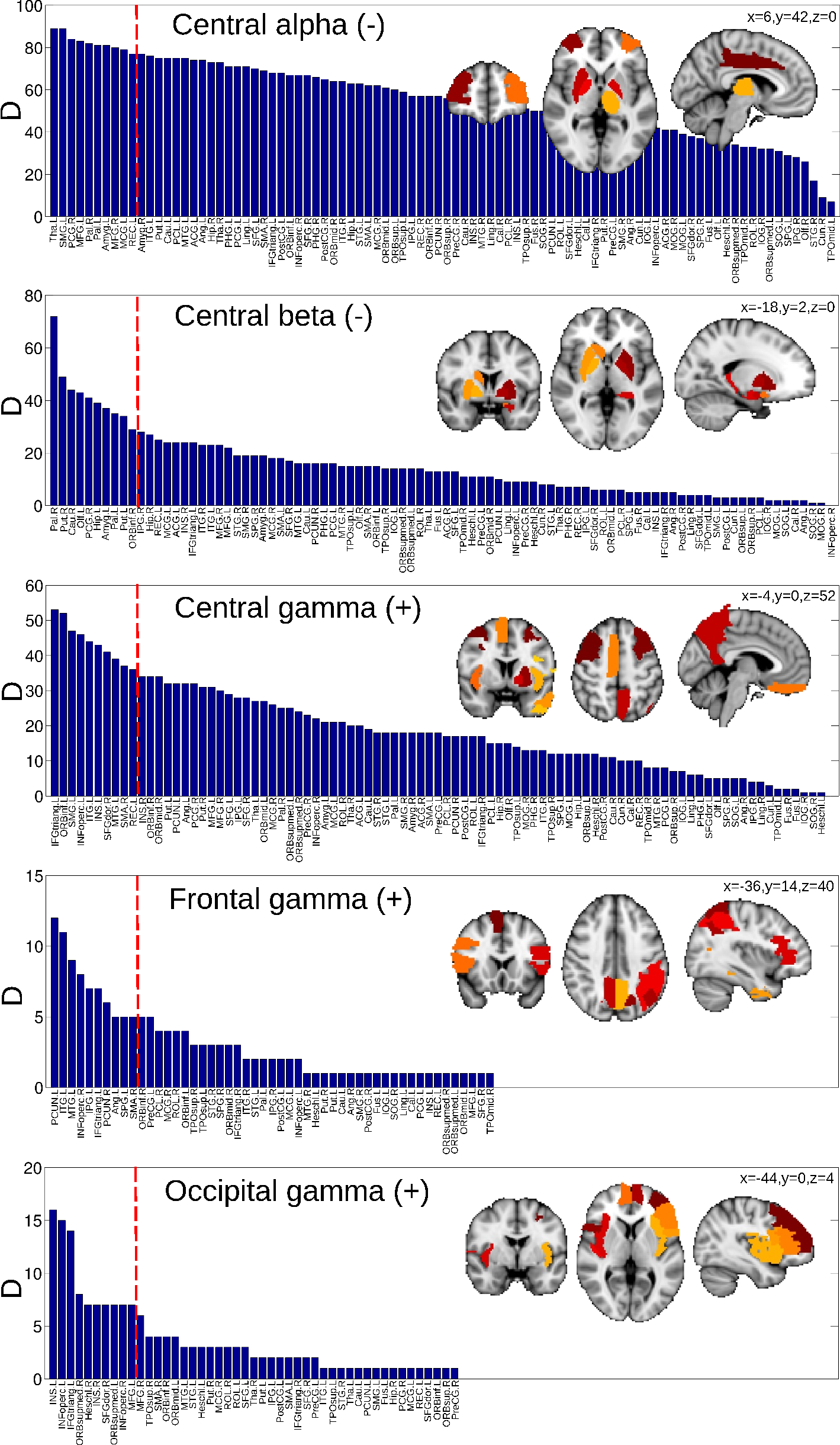}
\caption{Anatomical regions (or nodes) ranked according
 to the number of connections attached to them which correlate with EEG power fluctuations in different frequencies (equivalently, the
 degree $D$ of the nodes in a network in which each link represents BOLD connectivity correlating with EEG power). In the inset,
 the top 10 ranked regions are displayed overlaid on a standard MNI T1 template. Results 
 are for the group of awake subjects.}\label{fig6}
\end{figure}

\begin{figure}[htp]
\centering
\includegraphics[totalheight=0.8\textheight]{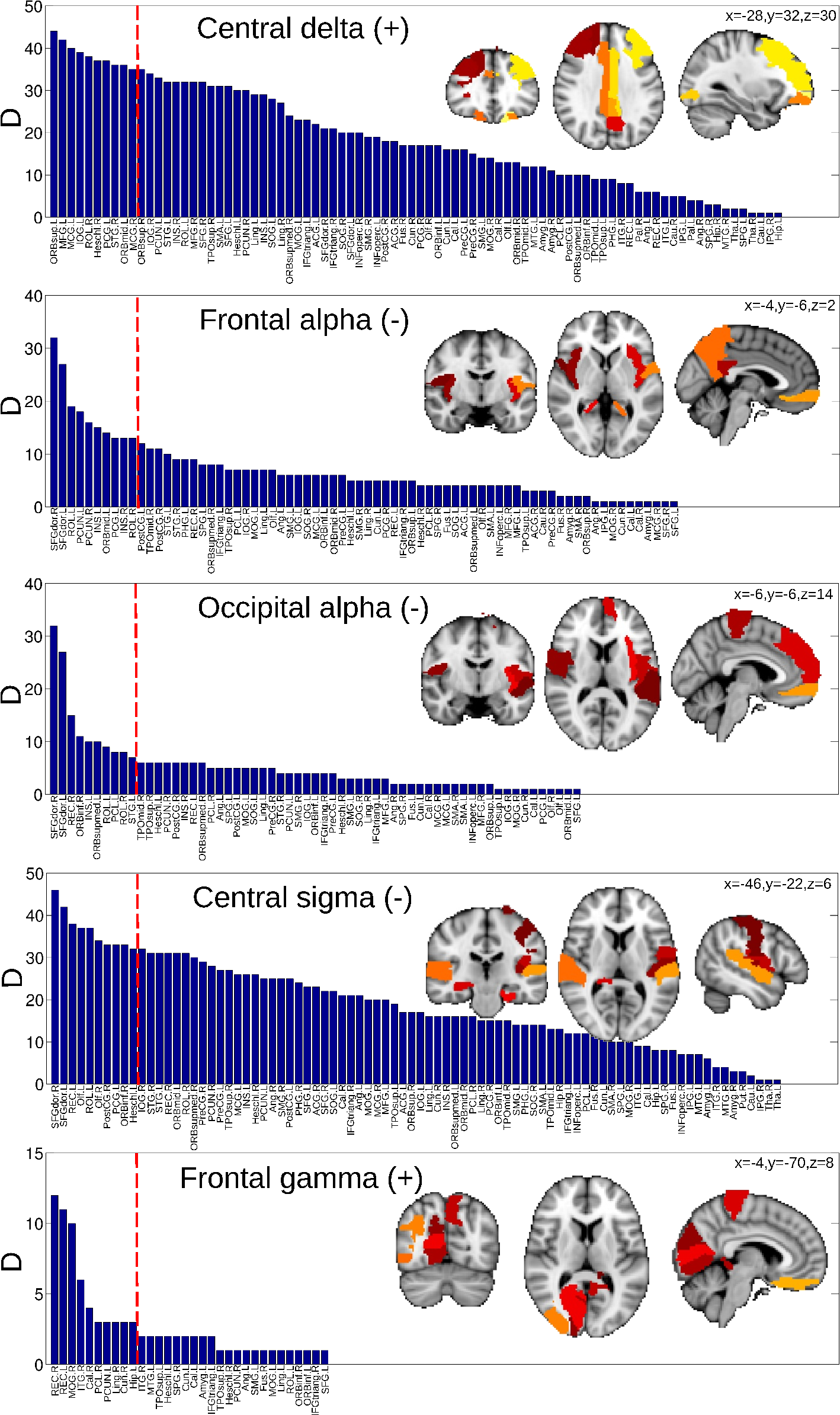}
\caption{Anatomical regions (or nodes) ranked according
 to the number of connections attached to them which correlate with EEG power fluctuations in different frequencies (equivalently, the
 degree $D$ of the nodes in a network in which each link represents BOLD connectivity correlating with EEG power). In the inset,
 the top 10 ranked regions are displayed overlaid on a standard MNI T1 template. Results 
 are for the group of subjects undergoing vigilance transitions to light sleep.}\label{fig7}
\end{figure}

After obtaining the degree for each region, they were ranked
in decreasing order. Results are shown in Fig. \ref{fig6} for the group of awake subjects and Fig. \ref{fig7} for the group of subjects
undergoing vigilance transitions to light sleep. In the insets, the top 10 regions with highest number of correlations with EEG power are displayed 
overlaid on a standard MNI T1 template
(lighter colors represent a higher number of correlations). For the group of awake subjects, the region with the highest number of negative correlations with 
central alpha was the thalamus. Other top-ranked regions were sub-cortical and bilateral frontal. 
For correlations with beta, highest ranked regions were also subcortical (pallidum, putamen and caudate nucleus). 
Top-ranked regions with positive functional connectivity correlations in the central gamma band were mostly frontal. The same was observed
for correlations with occipital gamma (with the inclusion of the bilateral insular cortex as the region with the highest degree). 
In correlations with frontal gamma, on the other gand, frontal regions were second to precuneus, temporal and parietal areas. 

For the group of subjects undergoing vigilance transitions, frontal  (superior and middle frontal gyri) and cingulate regions exhibited the 
largest number
of positive correlations with delta power.  The same was observed for negative correlations with frontal and occipital alpha (including also
and insular and precuneal cortex) and
central sigma (in all three cases, the top-ranked region was the dorsal part of the superior frontal gyrus). The top-ranked region
for positive correlations with frontal gamma was the gyrus rectus and occipital regions also were mostly affected.

\subsection{EEG power fluctuations and connectivity between different sets of brain regions}

Next, we quantified the number of connections modulated by EEG power between different sets of brain regions, in order to reveal 
frequency specific changes in communication between brain systems.

\begin{figure}[htp]
\centering
\includegraphics[totalheight=0.8\textheight]{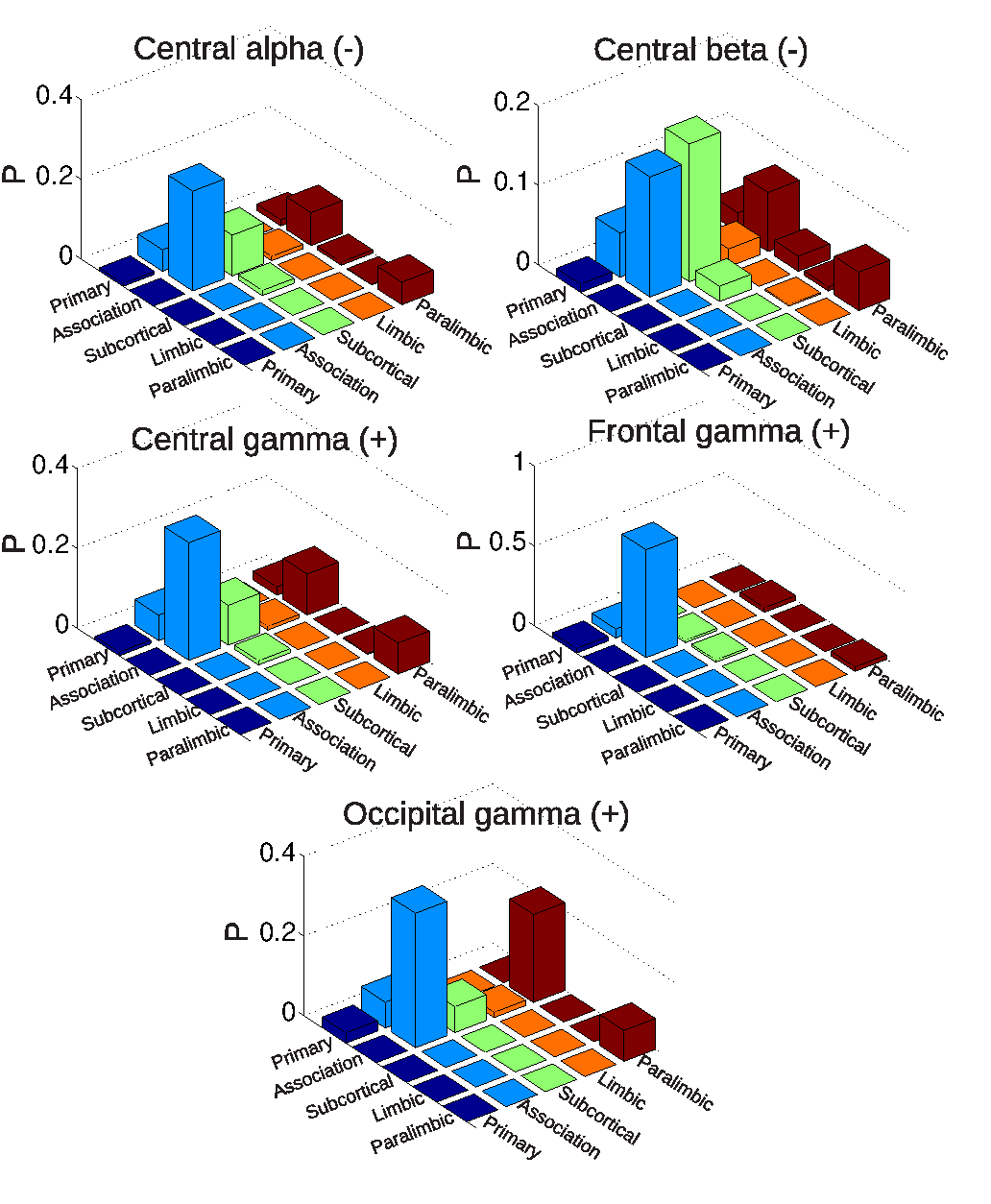}
\caption{Probability of finding connections between
 different systems (sensory, association, subcortical, limbic and paralimbic) which correlate either positively or negatively
 with spontaneous EEG power fluctuations.  Results are for the group of awake subjects.}\label{fig8}
\end{figure}

\begin{figure}[htp]
\centering
\includegraphics[totalheight=0.8\textheight]{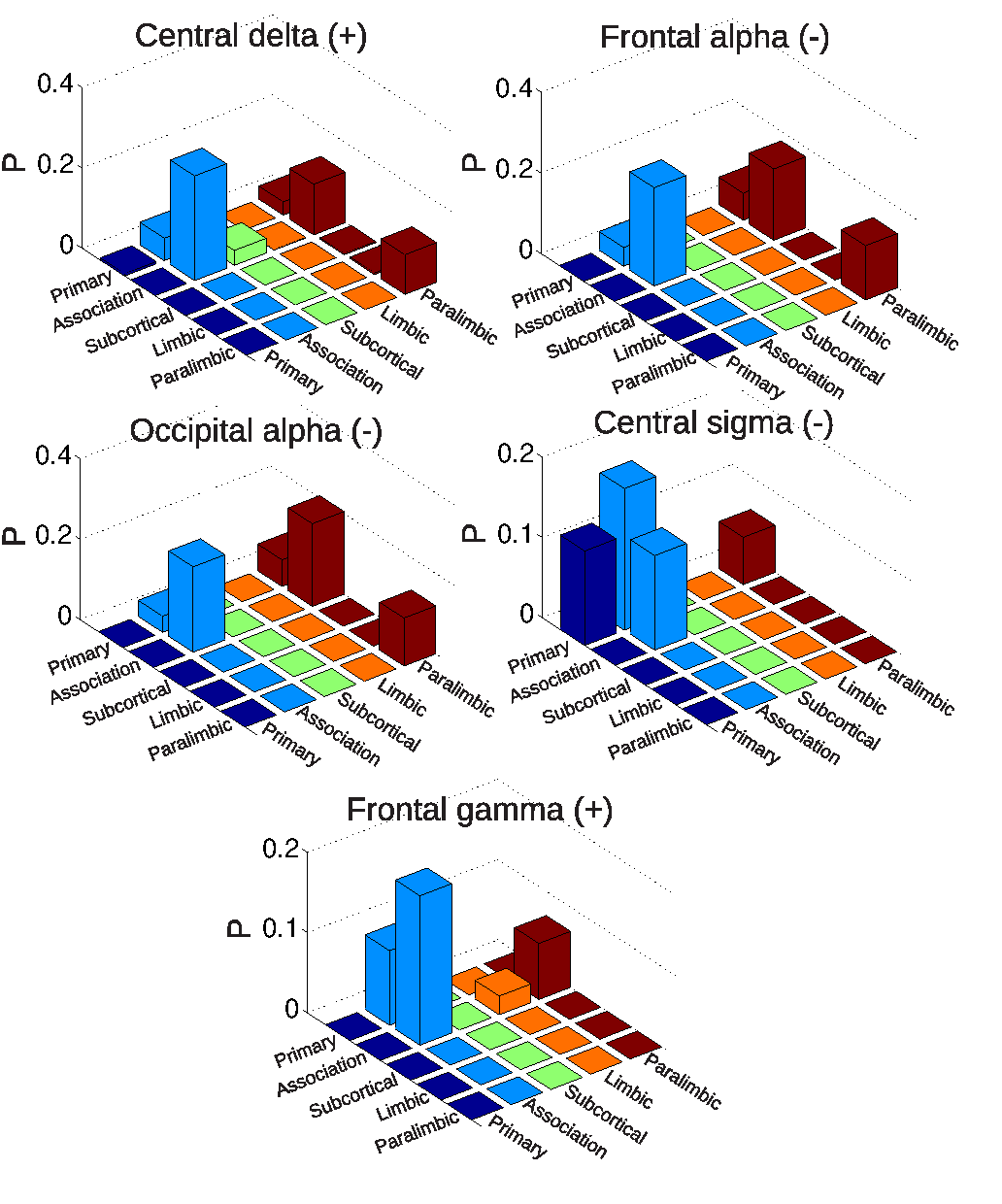}
\caption{Probability of finding connections between
 different systems (sensory, association, subcortical, limbic and paralimbic) which correlate either positively or negatively
 with spontaneous EEG power fluctuations.  Results are for the group of subjects undergoing vigilance transitions to light sleep.}\label{fig9}
\end{figure}

 For that purpose, a 
previously introduced classification of each region into five categories was followed, comprising primary sensory, association, subcortical, limbic and paralimbic \cite{achard}. 
The traffic between each pair of systems was computed as the normalized number of connections which covary negatively or positively with
spontaneous EEG power fluctuations. Results are shown in Fig. \ref{fig8} for the group of awake subjects and in Fig. \ref{fig9}
for the group of subjects undergoing vigilance transitions. For the first group, EEG alpha power mediates decreased
BOLD connectivity between the association and primary sensory cortices, as well as in subcortical and paralimbic systems. The effect of central
beta was similar but with a further reduction of connectivity between the association and subcortical and limbic systems. Positive
correlations with frontal gamma were confined inside the association systems only, whereas occipital and central gamma correlated with
BOLD connectivity between  association, primary, subcortical and paralimbic systems.  For the second group (wakefulness \& light sleep)
the most salient feature was a connectivity breakdown between primary and association systems and also
inside the primary sensory system itself, which was mediated by increased power in the sigma band. Frontal gamma positively influenced
few connections, located inside the association system and between the latter and primary, limbic and paralimbic systems.

\subsection{Correlation between EEG power and spontaneous graph metric fluctuations}

The observed correlations between EEG power fluctuations and BOLD connectivity suggest that graph theoretical measures of
network organization might also covary with EEG power changes. We studied this possibility for commonly studied graph metrics: clustering
coefficient, average path length, betweeness and small-worldness. 

\begin{figure}[htp]
\centering
\includegraphics[totalheight=0.5\textheight]{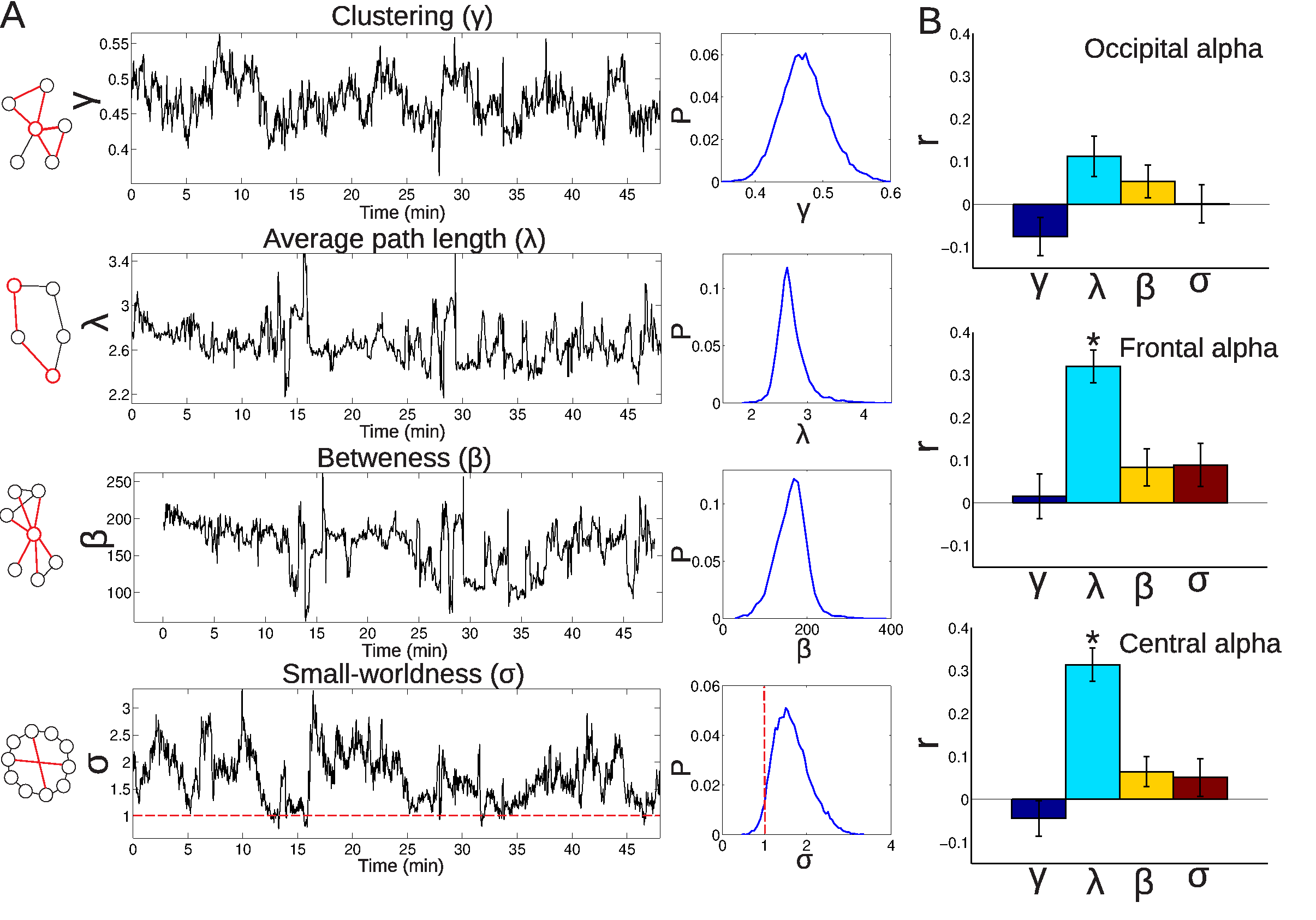}
\caption{Left: illustrations exemplifying the
 meaning of common graph metrics (clustering coefficient, average path length, betweeness and small worldness;
 for a detalied explanation, see the Methods section). Center: Examples showing the temporal 
 evolution of the graph metrics for a single subject. Right: Histograms (for all subjects) of the graph metric values. 
 B) Correlations between fluctuations in the graph metrics and EEG alpha power, averaged from different anatomical locations
 (*$p<0.05$, Bonferroni corrected, n=72).}\label{fig10}
\end{figure}

In Fig. \ref{fig10}A (left) an illustration exemplifying the
definitions provided in the Methods section is shown. Then an example of how the different graph metrics change during the duration
of a single subject measurement (Fig. \ref{fig10}A, center) is shown, together with histograms of graph metric values for all subjects (Fig. \ref{fig10}A, right). 
Note that while small-worldness ($\sigma$) 
is usually greater than 1 (the limit value at which networks are usually considered to be small-world \cite{humphries, vandenheuvel}), at certain times it goes below this value,
highlighting the fact that small-worldness of brain functional networks is inferred from average connectivity, and thus it can be lost at times.
In Fig. \ref{fig10}B the temporal correlations between fluctuations in graph metrics and occipital, frontal and central EEG alpha power are presented.
(alpha was the only frequency band with significant correlations). These correlations were positive between average path length and frontal-central alpha. 
An increased
average path length signals a more fragmented network, consistent with the widespread BOLD discoordination observed at times of high alpha power
(Fig. \ref{fig3}).

\section{Discussion}

In this paper we have studied how changes over time in BOLD
functional connectivity are linked to the transient local synchronization of scalp EEG rhythms. 
Our results reveal that increased EEG power in the gamma band facilitates long-range communication between brain regions, whereas slower frequencies
in the alpha and beta range are related to diminished functional connectivity. When studied using a graph theoretical approach, these results are manifest
in a positive correlation between power in the alpha band and average path length, an index proportional to the number of links 
which need to be crossed in order to get from one node to the other. An important implication of this observation is that network metrics are not
stationary: they widely fluctuate through the duration of an fMRI experiment. 

Our results are concordant with a recent study of primary visual
cortex connectivity modulation by spontaneous power fluctuations in posterior alpha \cite{scheeringa}, and represent an extension from regional 
and system-specific correlations
to the global relationship between EEG frequency power changes and  BOLD connectivity.   In the following, we discuss our results in the 
light of previous work and theories
about brain function, and we develop the most important implications of our results for future studies of brain functional connectivity.

\subsection{BOLD functional connectivity fluctuates at a time scale of minutes}

Our sliding window analysis revealed changes in functional connectivity at the scale of minutes. Such non-stationary connectivity 
 can be attributed to many factors or combination or factors: intrinsic
coupling and de-coupling of neural activity during rest, vigilance changes, spontaneous cognition, movement artifacts and changes in
cardiac or respiratory rates. A previous study demonstrated  temporal changes in connectivity (termed ``dynamical functional connectivity')
 in anesthetized macaques while ruling out motion
and spontaneous conscious cognition as the  origin of the observed BOLD connectivity changes \cite{hutchinson}. Our analysis
does not eliminate spontaneous cognition, rather, it links the connectivity changes to 
scalp rhythms involved in different brain states. Cardiac and motion time series did not
correlate with changes in functional connectivity, and respiratory time series only correlated with a minuscule fraction of the possible connections.
A large number of correlations with different EEG frequencies (discussed in detail below) -present even when motion, cardiac and respiratory time series
were taken into account as partial regressors- highlight the neural origin of the changes  in BOLD connectivity: the presence of specific neural 
oscillations is directly related to increased or decreased temporal locking of BOLD signals
from different cortical and subcortical areas.

 Furthermore, the standard deviation of sliding windowed BOLD connectivity, which quantifies the degree
of deviation from a stationary connectivity, displayed structured spatial variation: connections between neighboring regions 
 had a smaller standard deviation, while connections between regions from heterogeneous systems showed
larger deviations from constant connectivity. 

Finally, our analysis included two groups of subjects: one showing steady levels of vigilance (wakefulness) and other with transitions 
between wakefulness and light sleep. Given that the onset of sleep is characterized by changes in scalp EEG oscillations (a shift from fast towards slower frequencies \cite{AASM}),
a change in the relationship between BOLD connectivity fluctuations and EEG rhythms is expected in the group of subjects falling asleep. We observed
these changes: distributed positive temporal correlations between BOLD connectivity and EEG power in the slow delta band appeared, while positive
correlations with the fast gamma band mostly disappeared. Our results also suggest an electrophysiological correlate of changes
 in BOLD connectivity as previously reported at sleep onset \cite{spoormaker, enzo4}.

\subsection{Negative correlations with EEG power fluctuations}

We have shown that spontaneous increases in EEG alpha power (averaged over central channels) parallel decreased BOLD connectivity between
a large number of cortical and subcortical regions.
This would be consistent with the proposition of alpha as an ``idling rhythm'' which predominates
during relaxed, eyes closed rest \cite{pfurtscheller0} and with that of alpha as performing functional inhibition of regions not
relevant to task performance \cite{klimesch, jensen, scheeringa}. The transient BOLD connectivity decreases observed at times of large alpha amplitude could indicate
an active suppression of sensory input and the subsequent cortical processing \cite{worden}. While alpha is usually more prominent in occipital regions,
the blocking of oscillations in the same frequency range has been related to the onset and planning of activity  in sensory-motor and supplementary
 motor cortices \cite{pfurtscheller3} (these oscillations in the alpha range are termed ``mu'' or ``rolandic alpha rhythm'' \cite{pfurtscheller1}). Furthermore, previous EEG-fMRI studies
 demonstrated an inverse relationship between BOLD activation in a large, distributed
network of cortical areas \cite{laufs2003b, moosmann, goncalves} (overlapping with the default mode network \cite{laufs2003a}) and EEG alpha power. Overall, these results suggest that alpha
 suppression is a landmark of cortical activation, a view which is expanded by the present work by demonstrating a direct link with a transient increase
 in BOLD connectivity. 

The negative correlations obtained between BOLD connectivity and beta power are surprising, considering the traditional view of these faster rhythms 
as  a signal of increased mental activity, starting from their first observations by Hans Berger \cite{berger}. However, desynchronization of rolandic (peri-central)
 beta rhythms increases cortical excitability, favoring
a motor response \cite{deletis}. Beta rhythms over central regions appear in synchronized fashion after (but not during) the execution of a voluntary motor command \cite{pfurtscheller6, pfurtscheller5}.
 Because of this
 inverse relationship between rolandic beta rhythms and cortical excitability they are usually considered as idling
rhythms, which indicate a resting state of the sensory-motor cortex and related brain areas \cite{pfurtscheller2}. Furthermore, an inverse relationship between peri-central
beta rhythms and fMRI BOLD activation has been reported \cite{ritter}. Considered together, these results suggest that both peri-central beta and alpha rhythms are
 related to an idling state of the cortex, which is here related to a decreased BOLD functional connectivity. While in our study subjects did
not perform any explicit motor task, we hypothesize that during rest brain activity spontaneously recapitulates activity patterns related 
to the execution of diverse tasks, as well as stimuli perception - an hypothesis which has received strong experimental support from different
neuroimaging modalities \cite{ringach, smith,sadaghiani}. 

We also note that the connectivity of subcortical regions was most strongly affected by central and alpha rhythms. Since spontaneous thalamic
 BOLD activity has been shown to correlate positively with EEG alpha power, and activity from a large network of cortical areas shows an
 inverse correlation \cite{laufs2003a}, we can expect largely disrupted cortico-thalamic BOLD connectivity modulated by spontaneous changes in alpha power.
Subjects undergoing vigilance changes consistently displayed negatively correlated connectivity with alpha and sigma bands in frontal regions 
(bilateral superior frontal gyrus), which could be related to vigilance-related variability in correlation patterns with alpha power \cite{laufs2006}.

\subsection{Positive correlations with EEG power fluctuations}

In contrast to the slower frequencies, the fast gamma rhythm is almost universally related to the  binding of cortical areas active during 
different mental tasks. A long standing proposal in neuroscience assigns
to the gamma rhythm the role of aggregating anatomically distant brain regions into a coherent pattern of neural activity \cite{fries1}, i.e. 
a proposed solution to the ``binding problem'' \cite{malsburg, treisman}. Since the first pioneering studies showing long-range gamma synchronization during visual stimulation \cite{singer},
a wealth of experimental results has demonstrated the important role of synchronized activity in the gamma band during the execution of different cognitive tasks
(for reviews, see \cite{lee,herrmann,fries2} ). The ubiquity of gamma rhythms and their
apparently heterogeneous nature have led to the hypothesis that activity in the gamma band represents a  fundamental process subserving an elementary
 operation of cortical computation \cite{fries3}. Our results closely relate increased EEG gamma power 
(averaged from different topographical locations) to the transient coordination of BOLD signals between a large number of cortical and subcortical pairs of regions.
This result is consistent with these experiments and hypotheses if one accepts that fluctuations in BOLD activity and EEG 
power are not intrinsically different from changes elicited by stimulation or during task execution. This would be the case if spontaneous cognitive
operations  underlie resting state activity fluctuations, an hypothesis which has received experimental support \cite{andrews, shirer}. This
proposal, however, cannot be held as the
only origin of the aforementioned fluctuations, given the coordinated spatio-temporal activity observed in states of diminished conscious awareness, 
such as sleep \cite{prior, boly, horovitz, brodbeck}.
Finally, a close relationship between fluctuations in gamma power and BOLD connectivity is consistent with
repeated reports of correlations between the gamma band and BOLD activation \cite{logothetis, nir, lachaux}.	 .

Connectivity of frontal, precuneal and temporal regions with the rest of the brain was most strongly influenced by the increase of gamma power.
 EEG-fMRI studies
have shown that correlations between scalp EEG gamma and BOLD activity are located predominantly in frontal regions \cite{mantini}. 
Our study extends this
observation by demonstrating a direct relationship between increased gamma and BOLD connectivity of frontal regions with the rest of the brain. 
Evoked gamma activity in frontal regions appears to be a crucial feature of normal and healthy information processing in the cerebral cortex \cite{lee}.

Also, we observed positive correlations between increased delta power and BOLD connectivity, but only for the subjects undergoing vigilance
transitions to light sleep. Sleep onset is paralleled by large changes in the distribution of scalp EEG 
frequencies, with a decrease in the power of fast frequencies and increased power of slow oscillations (delta and theta bands) \cite{AASM}. 
These slow frequencies appear synchronously over the cerebral cortex but their nature is very different from that of the fast gamma rhythm: 
their low temporal complexity reflects the alternation between neural firing (``up'' state) and quiescence (``down'' state). Such lack of temporal
complexity, and therefore a diminished repertoire of possible neural states, is hypothesized to underlie loss of conscious 
awareness during sleep \cite{tononi1, tononi2, tononi3}. 
Our results suggest that BOLD over-synchronization follows increased activity in the delta band for subjects transitioning to light sleep.
Studies addressing the issue of temporal complexity (or temporal dependency) of BOLD signals across the human sleep cycle are needed
to reveal in fMRI recordings  the temporal properties of these slow neural oscillations.

\subsection{Temporal scales and correlations between BOLD connectivity and EEG power}

Given that changes in EEG power (for example, in beta or gamma bands) during or after task execution are usually transients in the sub-second temporal 
range, it is remarkable that correlations with the connectivity of the slow, lagged and relatively poorly sampled BOLD signal can be found.
However, a period with a particularly high level of activity could elicit a change in BOLD connectivity when all the short-lived EEG power changes
are considered together. This situation is analogous to that of the correlation between short periods of stable topographical configurations 
(EEG microstates \cite{koenig}) with specific BOLD RSN, which are likely driven by periods in which the presence of a given microstate overwhelmingly
exceeds that of the others \cite{britz}. It has been speculated and subsequently corroborated \cite{vandeville} that, for this to happen, the
 distribution of the EEG events has to
follow a scale-free distribution (or equivalently, have a $1/f$ spectrum): only then the temporal scale invariance allows the discovery of correlations
using a much slower imaging method, such as fMRI. These kind of distributions are ubiquitous in the power fluctuations of EEG, MEG
and electrocorticography (ECoG) recordings \cite{linkenkaer, miller, he,he2}, as well as in time series derived from cognitive
 and behavioral experiments \cite{gilden, shelhamer}.

\subsection{Hypersynchronization}

One particularly remarkable feature of dynamical functional connectivity is the existence of periods of extended BOLD synchronization \cite{hutchinson,fraiman}.
 In these
events, the BOLD signals from all the regions under consideration suddenly display coordinated activity lasting for a few seconds.
This \emph{hypersynchronization}
can hardly be attributed to spontaneous cognitive operations performed during rest, since it has been demonstrated in anesthetized macaques \cite{hutchinson}. 
A very
similar phenomenon has been described in a study reducing the BOLD signal to a point process and therefore allowing to quantify instantaneous 
connectivity as the temporal coincidence of points. A power law describes the number of points (or active voxels) coalescing into cortical
clusters at any given time, demonstrating the existence of scale-free avalanches of activity spreading throughout the complete cerebral cortex \cite{enzo3, enzo5}.
In the present work we have observed such hypersynchronous events and we have studied their electrophysiological correlates. Given
the widespread positive correlations between BOLD connectivity and EEG gamma power, we speculate that hypersynchrony (or, equivalently, activity
avalanches)  will be related to large, spontaneous increases in this frequency range. On the other hand, increased central alpha and beta power
correlate with the opposite phenomenon: a breakdown of large-scale BOLD functional connectivity. This is also manifest in the positive correlation
between alpha power and average path length, the average number of links which need to be crossed in order to get from one node to the other in the network.

\subsection{Graph metrics fluctuate over time}

We have demonstrated that during the time evolution of whole-brain functional networks, associated graph metrics (clustering coefficient,
average path length, betweeness and small-worldness) fluctuate widely. Given this result, a number of recent studies based
on the methodology of graph theory have to be re-interpreted. The reported value of the different graph metrics cannot be taken as a constant 
property of brain networks, instead, it has to be considered as an asymptotic property (i.e. the value one obtains during a long measurment) 
emerging after temporal averaging. There are two 
interesting immediate consequences of this observation. First, resting state studies applying graph theoretical methods need to be based
on long recordings, since short acquisition times will decrease the confidence on the graph metric estimates (chances will be higher
of computing them in a period in which they largely deviate from the mean). Second, the temporal evolution of graph metrics should 
be taken into account. For example, when comparing two populations using graph theoretical methods, equal values of the associated graph
metrics  could be obtained, yet their dynamical behavior could be completely different. Further investigations are needed to study this
and other possibilities of considering the dynamical evolution of functional networks over time.

\subsection{Implications for theories of conscious brain function}

Our results show that the onset of specific (fast) rhythms are paralleled by distributed binding of BOLD activity (i.e. increased
functional connectivity), whereas other
(slower) oscillations contribute to the inhibition such binding, decreasing the overall cortico-cortical and cortico-subcortical connectivity. The
fact that long-range functional connectivity in the human brain is unstable and fluctuates in a coordinated fashion with fast EEG rhythms 
likely reflects the dynamical nature of processes underlying the conscious state. For instance, an approach to consciousness focusing in its nature
as a \emph{process} (instead of a state or a capacity) emphasizes the presence of a \emph{dynamical core}, a continuously changing set of neuronal
groups strongly integrated together during hundreds of milliseconds and allowing differentiated responses, i.e. having a large neural complexity \cite{edelman2}.
Also, the dynamical core hypothesis resembles a dynamical instantiation of the global workspace theory, in which any sensory stimulation is postulated
to coalesce into a global pattern of activity in ``workspace neurons'', preventing conscious awareness of other concurrent stimuli \cite{dehaene}. This
sudden transition to a coherent and globally distributed pattern of activity has been modeled in graph-theoretical terms as the occurrence 
of transient links between pre-existing specialized processing modules, resulting in the formation of a giant connected component \cite{wallace}. 
Our results are reminiscent of this scenario, revealing that fast and temporally complex power oscillations in the gamma range are correlated 
with the appearance of long-range and large-scale network connectivity, while the alpha band positively correlates with a graph metric of network disconnection
(average path length).
 
While dynamical functional connectivity and its electrophysiological characteristics are suggestive of the processes postulated by the aforementioned
theories, further experimental tests are required in order to corroborate them as a correlate of conscious awareness (for example, studying whether
these interrelated, dynamical landmarks of brain activity are also prevalent during deep sleep or anesthesia).

\subsection{Conclusion}

Large efforts have been devoted to identifying the electrophysiological counterparts of the fMRI BOLD contrast. We have approached
this problem from a new perspective: studying whether the onset of band-specific scalp oscillations is related to increased (or decreased)
BOLD coherence, instead of directly relating it to changes in BOLD amplitude. Our results confirm the binding role of fast frequencies in the gamma range
and link slower ``idling'' rhythms with large-scale disconnection patterns, which are also quantified by correlation with adequate graph metrics.
The relationship found between EEG power fluctuations and dynamical BOLD functional connectivity leads us to conclude that this phenomenon is very likely
to be of neuronal origin, and thus it deserves even further investigation.

\section{Acknowledgments}

This work was funded by the Bundesministerium f\"ur Bildung und Forschung (grant 01 EV 0703) and the LOEWE
Neuronale Koordination Forschungsschwerpunkt Frankfurt (NeFF). The authors thank Torben E. Lund for providing a MATLAB implementation of the 
RETROICOR method, Sandra Anti, Ralf Deichmann and
Steffen Volz for extensive MRI support, and all subjects for their participation.

\clearpage

\thispagestyle{empty}
\pagestyle{empty}

\squeezetable 
\begin{table}[t]
\centering
\begin{tabular}{|c|c|c|c|c|}
\hline
\# (left-right) & Region name & Abbreviation &  System & Coordinates (left-right)\\ 
\hline
1-2 & Precentral gyrus &  PCG & Primary &  (37,    -6,    50) - (-42,    -4,   48)\\
3-4 & Superior frontal gyrus &  SFG & Association&   (17,    34,    41) - (-22,    36,    40)\\
5-6 & Superior frontal gyrus, orbital part&  ORBsup & Paralimbic&  (13,    48,   -17) - (-20,    47,   -17)\\
7-8 & Middle frontal gyrus&  MFG & Association&   (33,    34,    32) - (-36,    34,    32)\\
9-10 & Middle frontal gyrus, orbital&  ORBmid & Paralimbic&   (28,    53,   -14) - (-34,    50,   -13)\\
11-12 & Inferior frontal gyrus, opercular&  INFoperc & Paralimbic&   (46,    17,    19) - (-52,    13,    14)\\
13-14 & Inferior frontal gyrus, triangular&  INFtriang & Association &   (45,    32,    12) - (-48,   31,    10)\\
15-16 & Inferior frontal gyrus, orbital&  ORBinf & Paralimbic &  (36,    32,   -14) - (-39,    30,  -16)\\
17-18 & Rolandic operculum&  ROL & Association&   (48,    -4,    13) - (-50,    -8,    11)\\
19-20 & Supplementary motor area&  SMA & Association&   (4,     3,    60) - (-9,     8,    59)\\
21-22 & Olfactory cortex&  Olf & Primary &   (5,    16,   -14) - (-14,    13,   -15)\\
23-24 & Superior frontal gyrus, medial&  ORBsupmed & Paralimbic &   (4,    52,    28) - (-9,    51,    27)\\
25-26 & Superior frontal gyrus, dorsal&  SFGdor & Association &   (4,    52,   -11) - (-9,    55,   -11)\\
27-28 & Rectus gyrus &  REC  & Paralimbic &   (4,    34,   -21) - (-9,    37,   -22)\\
29-30 & Insula &  INS  & Paralimbic &   (34,     8,     0) - (-38,     7,    0)\\
31-32 & Anterior cingulate gyrus &  ACG  & Paralimbic &  (4,    38,    13) - (-8,    37,    10)\\
33-34 & Middle cingulate gyrus &  MCG  & Paralimbic&        (4,    -5,    38) - (-9,   -14,    39)\\
35-36 & Posterior cingulate gyrus &  PCG  & Paralimbic&    (4,   -40,    19) - (-8,   -41,    23)\\
37-38& Hippocampus &  Hip  & Limbic&   (24,   -20,   -11) - (-29,   -20,   -13)\\
39-40 & Parahippocampal gyrus &  PHG  & Paralimbic&   (21,   -15,   -22) - (-25,   -16,   -23)\\
41-42 & Amygdala &  Amyg  & Limbic&   (23,    1,   -19) - (-27,    -1,   -20)\\
43-44 & Calcarine cortex &  Cal  & Primary&   (12,   -73,     9) - (-11,   -79,     5)\\
45-46 & Cuneus &  Cun  & Association&    (10,   -79,    28) - (-9,   -79,    27)\\
47-48 & Lingual gyrus &  Ling  & Association&   (13,  -68,    -5) - (-18,   -69,    -6)\\
49-50 & Superior occipital gyrus &  SOG  & Association&   (20,   -78,    31) - (-19,   -84,    27)\\
51-52 & Middle occipital gyrus &  MOG  & Association&   (32,   -79,    19) - (-35,   -80,    15)\\
53-54 & Inferior occipital gyrus &  IOG  & Association &   (33,   -82,    -7) - (-40,   -78,    -9)\\
55-56 & Fusiform gyrus &  Fus  & Association&   (29,   -40,   -21) - (-34,   -41,   -22)\\
57-58 & Postcentral gyrus &  PostCG  & Primary &   (36,   -23,    51) - (-46,   -21,    47)\\
59-60 & Superior parietal gyrus &  SPG  & Association&   (22,   -56,    61) - (-27,   -58,    57)\\
61-62 & Inferior parietal gyrus &  IPG  & Association&   (42,   -44,    49) - (-46,   -44,    45)\\
63-64 & Supramarginal gyrus &  SMG  & Association&   (52,   -29,    33) - (-59,   -33,    28)\\
65-66 & Angular gyrus &  Ang  & Association&   (40,   -58,    39) - (-47,   -59,    33)\\
67-68 & Precuneus &  PCUN  & Association&   (6,   -54,    42) - (-10,   -54,    46)\\
69-70 & Paracentral lobule &  PCL  & Association&   (3,   -29,    66) - (-11,   -22,    68)\\
71-72 & Caudate &  Cau  & Subcortical&   (10,    12,     8) - (-15,    11,     7)\\
73-74 & Putamen &  Put  & Subcortical&   (23,     6,     1) - (-27,     4,     0)\\
75-76 & Pallidum &  Pal  & Subcortical&   (16,     0,    -1) - (-21,     0,    -2)\\
77-78 & Thalamus &  Tha  & Subcortical&   (8,  -17,     6) - (-14,   -18,     6)\\
79-80 & Heschl's gyrus &  Heschl  & Primary&   (39,   -16,     9) - (-47,   -18,     8)\\
81-82 & Superior temporal gyrus & STG  & Association&   (53,   -21,     6) - (-56,   -21,     5)\\
83-84 & Temporal pole (superior) & TPOsup  & Paralimbic&   (43,    15,  - -19) - (-44,    15,   -24)\\
85-86 & Middle temporal gyrus & MTG  & Association&   (53,   -37,    -2) - (-59,   -34,    -5)\\
87-88 & Temporal pole (middle) & TPOmid  & Paralimbic&  (40,    14,   -34) - (-40,    13,   -37)\\
89-90 & Inferior temporal gyrus & ITG  & Association&   (49,   -32,   -23) - (-53,   -29,   -26)\\
\hline
\end{tabular}
\caption{Region number, name, abbreviation, system membership (from \cite{achard}) and anatomical coordinates (for the center of mass of the 90 cortical and
 subcortical regions defined in the AAL template).}
\label{regionsAAL}
\end{table}

\clearpage


\end{document}